\documentclass[aoas,MSNbibl,nameyear,dvips]{arximspdf}
\usepackage{graphicx}
\doi{10.1214/11-AOAS533}
\volume{6}
\issue{3}
\pubyear{2012}
\firstpage{1327}
\lastpage{1347}

\makeatletter
\newproclaim{remark}{Remark}

\newcommand{\boldcapsigma}{\boldsymbol{\Sigma}}
\newcommand{\boldcappsi}{\boldsymbol{\Psi}}
\newcommand{\boldtheta}{\boldsymbol{\theta}}
\newcommand{\boldmu}{\boldsymbol{\mu}}
\makeatother

\begin{document}
\begin{frontmatter}

\title{Integrative Model-based clustering of microarray methylation
and expression data\thanksref{T1}}
\runtitle{Model-based clustering of microarray data}
\thankstext{T1}{Supported in part by NSF Grant DMS-08-05865.}

\begin{aug}
\author[A]{\fnms{Matthias} \snm{Kormaksson}\corref{}\ead[label=e1]{mk375@cornell.edu}},
\author[A]{\fnms{James G.} \snm{Booth}\ead[label=e2]{jb383@cornell.edu}},
\author[B]{\fnms{Maria~E.}~\snm{Figueroa}\ead[label=e3]{maf2049@med.cornell.edu}}
\and
\author[B]{\fnms{Ari} \snm{Melnick}\ead[label=e4]{amm2014@med.cornell.edu}}
\runauthor{Kormaksson, Booth, Figueroa and Melnick}
\affiliation{Cornell University, Cornell University, Weill Cornell
Medical College and~Weill Cornell Medical College}
\address[A]{M. Kormaksson \\
J. G. Booth\\
Department of Statistical Science\\
Cornell University\\
Ithaca, New York 14853 \\USA \\
\printead{e1}\\
\hphantom{\textsc{E-mail}:\ }\printead*{e2}} %adresu isvedimo
%komanda gale!
\address[B]{M. E. Figueroa \\
A. Melnick \\
Department of Medicine\\
Hematology Oncology Division\\
Weill Cornell Medical College\\
New York, New York 10065\\USA \\
\printead{e3}\\
\phantom{\textsc{E-mail}:\ }\printead*{e4}}
\end{aug}

% HISTORY:
\received{\smonth{12} \syear{2010}}
\revised{\smonth{11} \syear{2011}}

% ABSTRACT
%
\begin{abstract}
In many fields, researchers are interested in large and complex
biological processes. Two important examples are gene expression and
DNA methylation in genetics. One key problem is to identify aberrant
patterns of these processes and discover biologically distinct groups.
In this article we develop a model-based method for clustering such
data. The basis of our method involves the construction of a likelihood
for any given partition of the subjects. We introduce cluster specific
latent indicators that, along with some standard assumptions, impose a
specific mixture distribution on each cluster. Estimation is carried
out using the EM algorithm. The methods extend naturally to multiple
data types of a similar nature, which leads to an integrated analysis
over multiple data platforms, resulting in higher discriminating power.
\end{abstract}

% KEYWORDS
%

\end{frontmatter}
%
%s1 ###
%s1 #&#
\section{Introduction}

Epigenetics refers to the study of heritable characteristics not
explained by changes in the DNA sequence. The most studied epigenetic
alteration is cytosine (one of the four bases of DNA) methylation,
which involves the addition of a methyl group (a hydrocarbon group
occurring in many organic compounds) to the cytosine. Cytosine
methylation plays a fundamental role in epigenetically controlling gene
expression, and studies have shown that aberrant DNA methylation
patterning occurs in inflammatory diseases, aging, and is a hallmark of
cancer cells [\citet{Rodenhiser}; and \citet{Figueroa}]. \citet
{Figueroa} performed the first large-scale DNA methylation profiling
study in humans, where they hypothesized that DNA methylation is not
randomly distributed in cancer but rather is organized into highly
coordinated and well-defined patterns,\vadjust{\goodbreak} which reflect distinct
biological subtypes. Similar observations had already been made for
expression data [\citet{Golub}; \citet{Armstrong}]. Identifying such
biological subtypes through abnormal patterns is a~very important task,
as some of these malignancies are highly heterogeneous, presenting
major challenges for accurate clinical classification, risk
stratification and targeted therapy. The discovery of aberrant patterns
in subjects can identify tumors or disease subtypes and lead to a
better understanding of the underlying biological processes, which in
turn can guide the design of more specifically targeted therapies. Due
to the biological interaction between methylation and expression,
biologists hope to optimize the amount of biological information about
cancer malignancies by borrowing strength across both platforms. As an
example \citet{FigueroaPLOSone} showed that the integration of gene
expression and epigenetic platforms could be used to rescue genes that
were biologically relevant but had been missed by the individual
analyses of either platform separately.

In this article we propose a model-based approach to clustering such
high-dimensional microarray data. In particular, we build finite
mixture models that guide the clustering. These types of models have
been shown to be a principled statistical approach to practical issues
that can come up in clustering [\citet{McLachlan2}; \citet{Banfield};
\citet{Cheeseman}; \citeauthor{Fraley2} (\citeyear{Fraley2,Fraley1})]. The motivating application
is the cluster analysis of \citet{Figueroa}, which focused on patients
with Acute Myeloid Leukemia (AML). Both methylation and expression data
are available and we develop a clustering method that can be applied to
both data types separately. Furthermore, we extend our methodology to
facilitate an integrated cluster analysis of both data platforms
simultaneously. Although the methods are designed for these particular
applications, we expect that they can be applied to other types of
microarray data, such as ChIP-chip data.

A lot of attention has been given to classification based on gene
expression profiles and more recently based on methylation profiles.
\citet{Siegmund} give an overview and comparison of several clustering
methods on DNA methylation data. They point out that among biologists,
agglomerative hierarchical cluster analysis is popular. However, they
argue in favor of model-based clustering methods over nonparametric
approaches and propose a Bernoulli-lognormal model for the discovery of
novel disease subgroups. This model had previously been applied by
\citet{Ibrahim} to identify differentially expressed genes and profiles
that predict known disease classes. More recently, \citet{Houseman}
proposed a Recursively Partitioned Mixture Model algorithm (RPMM) for
clustering methylation data using beta mixture models [\citet{Ji}].
They proposed a beta mixture on the subjects and the objective was to
cluster subjects based on posterior class membership probabilities. The
RPMM approach is a model-based version of the HOPACH clustering
algorithm developed in \citet{Laan}.\vadjust{\goodbreak}

In high-dimensional data clustering is often performed on a smaller
subset of the variables. In fact, as pointed out in \citet{Tadesse},
using all variables in high-dimensional clustering analysis has proven
to give misleading results. There is some literature on the problem of
simultaneous clustering and variable selection [\citet{Friedman}; \citet
{Tadesse}; \citet{Kim}]. However, most statistical methods cluster the
data only after a suitable subset has been chosen. An example of such
practice is \citet{MaclachlanBeanPeel}, where the selection of a subset
involves choosing a significance threshold for the covariates. That is
also essentially what \citet{Houseman} and \citet{Figueroa} did, but
they selected a subset of the most variable DNA fragments. In this
paper we present an integrated model-based hierarchical clustering
algorithm that clusters samples based on multiple data types on the
most variable features. There is of course a~clear advantage of
automated variable selection methods such as in \citet{Tadesse}.
However, the implementation of such methods seems far from
straightforward and due to the popularity of hierarchical algorithms
among biologists [\citet{Kettenring} concluded that hierarchical
clustering was by far the most widely used form of clustering in the
scientific literature], there is a clear benefit in having a simple
hierarchical algorithm that can handle multiple data types.

The article is organized as follows. In Section \ref
{secdatadescription} we describe the features of the motivating data
set. In Section~\ref{secmodelspecif} we construct the model as a
mixture of Gaussian densities, which leads to a specific mixture
likelihood that serves as an objective function for clustering. We also
introduce individual specific parameters to account for subject to
subject variability within clusters (i.e., the array effect). In
Section~\ref{sectionhierarchicalclustering} we present two
model-based clustering algorithms. The first algorithm is a
hierarchical clustering algorithm that can be used to find a good
candidate partition. The second clustering algorithm is an iterative
algorithm that is designed to improve upon any initial partition. The
likelihood model can be applied to classification of new subjects and
in Section~\ref{secclassification} we describe a discriminant rule
for this purpose. In Section~\ref{secmultiplat} we extend the model
to account for multiple data platforms and in Sections \ref
{secanalysis} and~\ref{secotherapp} we apply the methods to real
data sets, which involve both methylation and expression data. We
conclude the article with a discussion in Section~\ref{secdiscussion}.

%s2 ###
%s2 #&#
\section{Motivating data} \label{secdatadescription}

\begin{figure}

\includegraphics{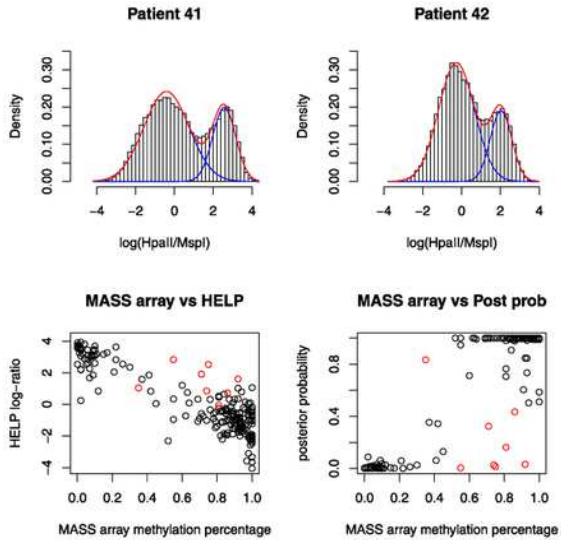}

\caption{Upper panel: A histogram of the log signal ratio, $\log
(\mathrm{HpaII}/\mathrm{MspI})$, for patients number $41$ and $42$ in the Erasmus data
set, along with two component Gaussian mixture fits. Lower panel: Left
graph shows HELP methylation values plotted against more accurate MASS
array methylation percentages. Right graph shows that the posterior
probabilities from a two component Gaussian mixture classifies probe
sets well into low and high methylation.}
\label{figurehistogrampatient234}\vspace*{-3pt}
\end{figure}

The Erasmus data were obtained from AML samples collected at Erasmus
University Medical Center (Rotterdam) between 1990 and 2008 and involve
DNA methylation and expression profiles of $344$ patient specimens. For
each specimen it was confirmed that $>$90\% of the cells were blasts
(leukemic cells). Description of the sample processing can be found in
\citet{Valk} and data sets are available in GEO,
\url{http://www.ncbi.nlm.nih.gov/geo/},\vadjust{\goodbreak} with accession numbers GSE18700 for
the methylation data and GSE$6891$ for the expression data. The gene
expression profiles of the AML samples were determined using
oligonucleotide microarrays (Affymetrix U133Plus2.0 GeneChips) and were
normalized using the rma normalization method of \citet{RMA}. The
processed data involved 54,675 probe sets and demonstrated a right
skewed distribution of the expression profiles for each subject; see
Supplementary Figure 1 in the supplemental article [\citet
{Kormaksson}]. The methylation profiles of the AML samples were
determined using high density oligonucleotide genomic HELP arrays from
NimbleGen Systems that cover 25,626 probe sets at gene promoters, as
well as at imprinted genes. Briefly, genomic DNA is isolated and
digested by the enzymes HpaII and MspI, separately. While HpaII is only
able to cut the DNA at its unmethylated recognition motif (genomic
sequence 5$'$-CCGG-3$'$), MspI cuts the DNA at any HpaII site whether
methylated or unmethylated. Following PCR, the HpaII and MspI digestion
products are labeled with different fluorophores and then cohybridized
on the microarray. This results in two average signal intensities that
measure the relative abundances (in a population of cells) of MspI and
HpaII at each probe set. The data are preprocessed and normalized using
the analytical pipeline of \citet{Reid} and the final quantity of
interest is $\log(\mathrm{HpaII}/\mathrm{MspI})$. Note that although theoretically
HpaII should always be less than MspI, complex technical aspects that
arise during the preparation and hybridization of these samples may
result in an enrichment of the HpaII signal over that of MspI.
Therefore, the log-ratio does not have a one-to-one correspondence with
percent methylation at a given probe set but rather provides a relative
methylation value that correlates with actual percentage value (see
lower left panel of Figure~\ref{figurehistogrampatient234}).
%
%f1 ###
%f1 #&#

In what follows, notation will be based on the HELP methylation data.
However, if we abstract away from this particular application, the
terminology can be adapted to other microarray data such as gene
expression. Let $y_{ij}$ denote the continuous response $\log
(\mathrm{HpaII}_{ij}/\mathrm{MspI}_{ij})$ for subject $i=1,\ldots,n$, and probe set
$j=1,\ldots,G$. Lower values of $y_{ij}$ indicate that probe set $j$
has high levels of methylation (in a population of cells) for subject
$i$, whereas higher values indicate low levels of methylation. In the
upper panel of Figure~\ref{figurehistogrampatient234} we see bimodal
histograms of the methylation profiles for two patients in the AML data
set along with two component Gaussian mixture fits. In Supplementary
Figure 2 of the supplemental article [\citet{Kormaksson}] we see
density profiles for all $344$ samples stratified by clusters. There is
a large microarray effect in the methylation data, but we observe that
all profiles are either skewed or exhibit a bimodal behavior. The lower
left panel of Figure~\ref{figurehistogrampatient234} shows how the
HELP assay correlates with methylation percentages obtained using the
more accurate, but much more expensive, quantitative single locus DNA
methylation validation MASS Array [see \citet{Figueroa}]. It is clear
that the HELP values are forming two clusters of relatively low or high
methylation levels with some noise in the percentage range $[20\%,80\%
]$. This apparent dichotomization inspires modeling each individual
profile, $\mathbf{y}_i=(y_{i1},\ldots,y_{iG})'$, with a two component
mixture distribution and normality is assumed for each component due to
its flexibility and ease of implementation. We know of no biological
mechanism that would imply normality, however, the assumption gives
consistent and reasonable fits of the individual methylation profiles
(see upper panel of Figure~\ref{figurehistogrampatient234}).

%s3 ###
%s3 #&#
\section{Model specification} \label{secmodelspecif}

By dichotomizing the methylation process we can cluster the probe sets
into high or low methylation for each patient $i$ by applying a two
component Gaussian mixture model. Let $\mathcal{C}$ denote the true
partition of the subject set, $[n] = \{1,\ldots,n\}$. We assume that on
any given probe set $j$, all subjects sharing a cluster $c \in\mathcal
{C}$ have the same relative methylation status (high or low),
and introduce for each cluster $c$ a single latent indicator vector,
$\mathbf{w}_c=(w_{c1},\ldots,w_{cG})'$, with
%
%e1 ###
%e1 #&#
\begin{equation}\qquad
w_{cj} =
\cases{\displaystyle  1 ,&\quad  if $j$ has high methylation for all
subjects in cluster $c$,\cr\displaystyle
0 ,&\quad  if $j$ has low methylation
for all subjects in cluster $c$.
}
  \label{methylationindicator}
\end{equation}
It is well known that methylation does exhibit biological variability
from individual to individual. However, it is biologically reasonable\vadjust{\goodbreak}
to expect consistency in relative methylation patterns for patients
that share the same disease subtype. Define $\boldtheta_i=(\mu
_{1i},\sigma_{1i}^2,\mu_{2i},\sigma_{2i}^2)'$ and assume that the
observed data, $\mathbf{y} =(\mathbf{y}_1',\ldots,\mathbf{y}_n')'$,
given the unobserved methylation indicators, $\mathbf{w} =(\mathbf
{w}_1',\ldots,\mathbf{w}_K')'$ ($K$ being the number of clusters),
arise from the following density:
%
%e2 ###
%e2 #&#
\begin{equation}
f(\mathbf{y}|\mathbf{w},\boldtheta) = \prod_{c \in\mathcal{C}}
\prod_{i \in c} f(\mathbf{y}_i|\mathbf{w}_c,\boldtheta_i), \label
{eqnclasslik}
\end{equation}
with
%
%e3 ###
%e3 #&#
\begin{equation}
f(\mathbf{y}_i|\mathbf{w}_c,\boldtheta_i) = \prod_{j=1}^G \phi
(y_{ij}|\mu_{1i},\sigma_{1i}^2)^{w_{cj}} \phi(y_{ij}|\mu
_{2i},\sigma_{2i}^2)^{1-w_{cj}}, \label{eqnsinglepatientdensity}
\end{equation}
where $\phi$ denotes the normal density and $\boldtheta=(\boldtheta
_1',\ldots,\boldtheta_n')'$. We refer to the density in (\ref
{eqnclasslik}) as the classification likelihood of the observed data
[\citet{Scott}; \citet{Symons}; \citet{Banfield}] and assume $\mu_{1i}
< \mu_{2i}$ for all $i$. We interpret $\boldtheta_i$ as the
individual specific means and variances of the high and low methylation
probe sets, respectively. Note that this setup is different from the
usual model-based clustering setup where we have cluster specific
parameters only. However, due to array effects, it is reasonable and in
fact necessary to require different parameters for different subjects.
In the upper panel of Figure~\ref{figurehistogrampatient234} we see
histograms and fits for two patients that both have chromosomal
inversions at chromosome $16$, inv($16$) (inversions refer to when
genetic material from a chromosome breaks apart and then, during the
repair process, it is re-inserted back but the genetic sequence is
inverted from its original sense). These two patients cluster together
under various clustering algorithms, including the model-based
algorithm presented below. However, the two distributions are clearly
not equal.

We put a Bernoulli prior on the latent methylation indicators in (\ref
{methylationindicator}):
%
%e4 ###
%e4 #&#
\begin{equation}
f(\mathbf{w}) = \prod_{c \in\mathcal{C}} \prod_{j=1}^G \pi
_{1c}^{w_{cj}}\pi_{0c}^{1-w_{cj}}, \qquad\pi_{0c} + \pi_{1c} = 1,
\label{densityw}
\end{equation}
where $\pi_{1c}$ and $\pi_{0c}$ denote the proportions of probe sets
that have high and low methylation, respectively, in cluster $c$. From
(\ref{eqnclasslik}) and~(\ref{densityw}) it is clear that the
complete data density is
%
%e5 ###
%e5 #&#
\begin{equation}\label{completelik}
 f(\mathbf{y},\mathbf{w}) = \prod_{c \in\mathcal{C}} \prod_{j=1}^G
 \biggl(\pi_{1c}\prod_{i\in c}\phi(y_{ij}|\mu_{1i},\sigma_{1i}^2)
 \biggr)^{w_{cj}}  \biggl(\pi_{0c}\prod_{i \in c}\phi(y_{ij}|\mu
_{2i},\sigma_{2i}^2)  \biggr)^{1-w_{cj}},\hspace*{-30pt}
\end{equation}
and if we integrate out the latent variable $\mathbf{w,}$ we arrive
at the marginal likelihood
%
%e6 ###
%e6 #&#
\begin{equation}
L_\mathcal{C}(\boldcappsi) = \prod_{c \in\mathcal{C}} \prod
_{j=1}^G \biggl( \pi_{1c} \prod_{i \in c} \phi(y_{ij}|\mu_{1i},\sigma
_{1i}^2) + \pi_{0c} \prod_{i \in c} \phi(y_{ij}|\mu_{2i},\sigma
_{2i}^2)  \biggr), \label{eqnmarginallikelihoodalldata}
\end{equation}
where $\boldcappsi= \{(\pi_{1c})_{c \in\mathcal{C}},(\mu
_{1i},\sigma_{1i}^2,\mu_{2i},\sigma_{2i}^2)_i \}$ denotes the set of
parameters. This likelihood can be used as an objective function for
determining the goodness of different partitions and the maximization
of~(\ref{eqnmarginallikelihoodalldata}) is carried out with the EM
algorithm of \citet{Dempster}. Note that $L_\mathcal{C}$ can be
written as a product, $\prod_{c \in\mathcal{C}} L_c$, where $L_c$
denotes the likelihood contribution of cluster $c$. Thus, maximizing
$L_\mathcal{C}$ can be achieved by maximizing $L_c$ independently for
all $c \in\mathcal{C}$. Details of the maximization algorithm are
provided in Supplementary Appendix B of the supplemental article [\citet
{Kormaksson}].\looseness=-1

%re1 #&#
\begin{remark} \label{remark1}
The premise of the clustering algorithm presented in Section~\ref
{sectionhierarchicalclustering} is to cluster subjects together that
have similar methylation patterns. Similarities across the genome in
the posterior probabilities of high/low methylation guide which
subjects are clustered together and, thus, if the posterior probability
predictions reflect the data well, the clustering algorithm should
perform well. In the lower right panel of Figure \ref
{figurehistogrampatient234} we see that the posterior probabilities
of high methylation fit very well with the actual percentage values.
\end{remark}

%
%re2 #&#
\begin{remark} \label{remark2}
When we allow for unequal variances $\sigma_{1i}^2 \neq\sigma
_{2i}^2$, the likelihood in~(\ref{eqnmarginallikelihoodalldata}) is
unbounded and does not have a global maximum. This can be seen by
setting one of the means equal to one of the data points, say, $\mu
_{1i}=y_{ij}$, for some $i,j$. Then the likelihood approaches infinity
as $\sigma_{1i}^2 \rightarrow0+$. However, \citet{McLachlan}, using
the results of \citet{Kiefer}, point out that, even though the
likelihood is unbounded, there still exists a~consistent and
asymptotically efficient local maximizer in the interior of the
parameter space. They recommend running the EM algorithm from several
different starting values, dismissing any spurious solution (on its way
to infinity), and picking the parameter values that lead to the largest
likelihood value.\looseness=-1
\end{remark}

%
%re3 #&#
\begin{remark}
Note that the likelihood in~(\ref{eqnmarginallikelihoodalldata}) is
identifiable except for the standard and unavoidable label switching
problem in finite mixture models [see, e.g., \citet{McLachlan}].
Furthermore, there exists a sequence of consistent local maximizers, as
$G \rightarrow\infty$. This becomes more evident if one recognizes
that the expression for a single cluster $c$ can be written as
a~multivariate normal mixture
%
%e7 #&#
\[
L_c = \prod_{j=1}^G \bigl( \pi_{1c} \phi(\mathbf{y}_{cj}|\boldmu
_{1c},\boldcapsigma_{1c}) + \pi_{0c} \phi(\mathbf{y}_{cj}|\boldmu
_{2c},\boldcapsigma_{2c})  \bigr),
\]
where $\mathbf{y}_{cj} = (y_{1j},\ldots,y_{n_cj})'$, $\mu_{kc} = (\mu
_{k1},\ldots,\mu_{kn_c})'$ and $\boldcapsigma_{kc}=
\operatorname{diag}(\sigma_{ki}^2)_{i=1}^{n_c}$, $k=1,2$ (assuming for convenience
that $i=1,\ldots,n_c$ are the members of cluster $c$).
Standard theory thus applies [see \citet{McLachlan}].
\end{remark}

%
%re4 #&#
\begin{remark}
The correlation structure of high-dimensional microarray data is
complicated and hard to model. Thus, we assume independence\vadjust{\goodbreak} across
variables in the likelihood~(\ref{eqnmarginallikelihoodalldata}) even
though it may not be the absolutely correct model. However, we can view
(\ref{eqnmarginallikelihoodalldata}) as a composite likelihood [see
\citet{Lindsay}] which yields consistent parameter estimates but with a
potential loss of efficiency. The correlations observed in microarray
data are usually mild and involve only a few and relatively small
groups of genes that have moderate or high within-group correlations.
In Supplementary Appendix A of the supplemental article [\citet
{Kormaksson}] we perform a simulation study to get a sense of how
robust our algorithm is to this independence assumption. The results
indicate that with a sparse overall correlation structure in which
genes tend to group into many small clusters with moderate to high
within-group correlations, our algorithm is not affected by assuming
independence across variables. However, there is some indication that
with larger groups of genes with very extreme within-group correlations
the algorithm will break down. In microarray data such extreme
correlation structures are not to be expected on a global scale and,
therefore, we believe that the independence assumption is quite reasonable.
\end{remark}

%s4 ###
%s4 #&#
\section{Model-based clustering} \label{sectionhierarchicalclustering}

Our clustering criterion involves finding the partition that gives the
highest maximized likelihood $L_\mathcal{C}$ as given in (\ref
{eqnmarginallikelihoodalldata}). This provides us with a model
selector, as we can compare the maximized likelihoods of any two
candidate partitions. In theory we would like to maximize $L_\mathcal
{C}$ with respect to all possible partitions of $[n]$ and simply pick
the one resulting in the highest likelihood. However, as this is
impossible for even moderately large $n$, we propose two clustering
algorithms. In Section~\ref{subsechierarch} we propose a simple
hierarchical clustering algorithm, and in Section \ref
{seciterativeclustering} we propose an iterative algorithm that is
designed to improve upon any initial partition.

%s4.1 ###
%s4.1 #&#
\subsection{Hierarchical clustering algorithm} \label{subsechierarch}
In this subsection we describe a~simple hierarchical algorithm that
attempts to find the partition that maximizes $L_\mathcal{C}$ as given
in~(\ref{eqnmarginallikelihoodalldata}). \citet{heard} used a~similar
approach, but they constructed a hierarchical Bayesian clustering
algorithm that seeks the clustering leading to the maximum marginal
posterior probability. The algorithm can be summarized in the following steps:
\begin{longlist}[3.]
\item[1.] We start with the partition where each subject represents its own
cluster, $\mathcal{C}_1=\{ \{1\},\ldots,\{n\}\}$, and calculate the
maximized likelihood, $L_{\mathcal{C}_1}$. Note that this likelihood
can be written as a product $L_{\mathcal{C}_1}=\prod_i L_{\{i\}}$
and, thus, the first step involves maximizing $L_{\{i\}}$ for each
$i=1,\ldots,n$. This is achieved by fitting a two-component Gaussian
mixture to each of the $n$ individual profiles. As mentioned in Remark
\ref{remark2}, each fit can be obtained by using the EM algorithm
starting from several different initial values and finding a local
maximum. It is important that the user verifies these initial
individual fits before\vadjust{\goodbreak} proceeding with the hierarchical algorithm. For
example, by going through the $344$ methylation profile fits of the
Erasmus data, one by one, we observe pleasing fits. The upper panel of
Figure~\ref{figurehistogrampatient234} gives examples of two such
profile fits.
\item[2.] Next we merge the two subjects that leads to the highest value of
$L_\mathcal{C}$ and denote the maximized likelihood value by
$L_{\mathcal{C}_2}$. Note that there are ${n \choose2}$ many ways of
merging two subjects at this step. However, since we already obtained
fits for $L_{\{i\}}$, $i=1,\ldots,n$, at Step 1, we only need to
maximize $L_{\{i,i'\}}$, for all pairs $(i,i')$ and find the pair that maximizes
%
%e8 #&#
\[
\ell_{\mathcal{C}_2} = \ell_{\mathcal{C}_1} - \bigl(\ell_{\{i\}} + \ell
_{\{i'\}}\bigr) + \ell_{\{i,i'\}},
\]
where $\ell$ denotes the loglikelihood. Even though we are applying
several EM algorithms, the complexity of each algorithm is low since it
only involves two subjects at a time.
\item[3.] We continue merging clusters under this maximum likelihood
criteria, at each step making note of the maximized likelihood, until
we are left with one cluster containing all $n$ subjects, $\mathcal
{C}_n=[n]$. Among the $n$ partitions that are obtained, we pick the
partition that has the highest value of $L_\mathcal{C}$. Note that the
likelihood value may either increase or decrease at each step. This
provides us with a method that automatically determines the number of clusters.
\end{longlist}
It is our experience that the individual parameter estimates $(\mu
_{1i},\sigma_{1i}^2,\mu_{2i},\sigma_{2i}^2)_i$ do not change much at
each merging step of the hierarchical algorithm. Thus, if the initial
estimates provide good fits for all the individual profiles, the
algorithm can be expected to perform well. Furthermore, by using the
individual parameter estimates at a previous merging step as initial
values at the next step, each EM algorithm converges very quickly,
which is essential since the total number of EM algorithms that are
conducted is of the order $O(n^2)$. For the data sets that we consider
in this article, the hierarchical algorithm takes anywhere from a
couple of minutes to run, for the smallest data set in Section \ref
{subsecendometrial} ($n=14$), up to a couple of hours for the Erasmus
high-dimensional data set of Section~\ref{secclusteringresults}
($n=344$), using a regular laptop. However, it should be noted that our
R code is neither optimized nor precompiled to a lower level
programming language at this stage.

%s4.2 ###
%s4.2 #&#
\subsection{Iterative clustering algorithm} \label{seciterativeclustering}

The hierarchical algorithm results in a partition that serves as a good
initial candidate for the true partition. In this subsection we present
an iterative algorithm that is designed to improve upon any initial
partition. We introduce cluster membership indicators for the subjects
in order to develop an EM algorithm for clustering subjects under the
assumption of a fixed number of clusters. Define for each subject
$i=1,\ldots,n$ and cluster $c \in\mathcal{C}$
%
%e9 #&#
\[
X_{ic} =
\cases{\displaystyle 1 ,&\quad  if subject $i$ is in cluster $c$,\cr\displaystyle
0 ,&\quad  otherwise,
}
\]
and let $\mathbf{X}_i = (X_{ic})_{c \in\mathcal{C}}$. Assume
$\mathbf{X}_1,\ldots,\mathbf{X}_n$ are i.i.d. Multinom$\{1,\mathbf
{p}=(p_c)_{c \in\mathcal{C}}\}$, so the density of $\mathbf{X} =
(\mathbf{X}_1',\ldots,\mathbf{X}_n')'$ is
%
%e7 ###
%e10 #&#
\begin{equation}
f(\mathbf{X}) = \prod_{i=1}^n \prod_{c \in\mathcal{C}}
p_c^{X_{ic}}, \qquad\sum_{c \in\mathcal{C}} p_c = 1. \label
{eqnclustermembership}
\end{equation}
These cluster membership indicators fully define the partition
$\mathcal{C}$ and we note that the classification likelihood in (\ref
{eqnclasslik}) can be written as
%
%e8 ###
%e11 #&#
\begin{equation}
f(\mathbf{y}|\mathbf{X}) = \prod_{c \in\mathcal{C}} \prod_{i=1}^n
f(\mathbf{y}_i | \mathbf{w}_c,\boldtheta_i)^{X_{ic}}. \label
{eqntwowaylikelihood}
\end{equation}
Multiplying~(\ref{eqnclustermembership}) and (\ref
{eqntwowaylikelihood}) together and integrating out $\mathbf{X}$, we
arrive at the marginal likelihood
%
%e9 ###
%e12 #&#
\begin{equation}
f(\mathbf{y};\boldcappsi) = \prod_{i=1}^n \sum_{c \in\mathcal{C}}
p_c f(\mathbf{y}_i|\mathbf{w}_c,\boldtheta_i), \label{marginalfixedlik}
\end{equation}
where $\boldcappsi= \{(p_c)_{c \in\mathcal{C}},(\mathbf{w}_c)_{c
\in\mathcal{C}},\boldtheta_i=(\mu_{1i},\sigma_{1i}^2,\mu
_{2i},\sigma_{2i}^2)_i \}$ involves both the continuous parameters and
the discrete indicators, $\mathbf{w}$, which we now assume are fixed.
We make this assumption because if $\mathbf{w}$ is random as in (\ref
{densityw}), the joint posterior distribution of $(\mathbf{w},\mathbf
{X})$ is highly intractable and an EM algorithm based on (\ref
{eqntwowaylikelihood}) would be problematic.

The likelihood in~(\ref{marginalfixedlik}) is that of a finite mixture
model and can be maximized using the EM algorithm. We detail the
maximization procedure in Supplementary Appendix B of the supplemental
article [\citet{Kormaksson}]. In short, let $\mathbf{X}^{(0)}$ denote
the clustering labels corresponding to a~candidate partition. Using
$\mathbf{X}^{(0)}$ as an initial partition, we run an EM algorithm
that converges to a local maximum of~(\ref{marginalfixedlik}). Once
the mixture model has been fitted, a probabilistic clustering of the
subjects can be obtained through the fitted posterior expectations of
cluster membership for the subjects, $(E[X_{ic}|\mathbf{y}])_{i,c}$
[see \citet{McLachlan}]. Essentially, a subject will be assigned to the
cluster to which it has the highest estimated posterior probability of
belonging. We have found empirically that the derived partition not
only results in a higher value of~(\ref{marginalfixedlik}) but also in
the objective likelihood~(\ref{eqnmarginallikelihoodalldata}), but we
do not have a theoretical justification for this. A good clustering
strategy is to come up with a few candidate partitions, with varying
numbers of clusters, and run the EM algorithm using these partitions as
initial partitions. Each resulting partition will be a local maximum of
(\ref{marginalfixedlik}), but we choose the partition with the highest
value of the original objective function (\ref
{eqnmarginallikelihoodalldata}). Good initial partitions can be found
by running the hierarchical algorithm of Section \ref
{subsechierarch} or applying one of the more standard clustering
algorithms.\looseness=-1

%s5 ###
%s5 #&#
\section{Classification} \label{secclassification}

The construction of a likelihood for any given partition of the
subjects also provides a powerful tool for classification. Assume we
have methylation data on $n$\vadjust{\goodbreak} subjects and we know which class each
subject belongs to, that is, we know the true $\mathcal{C}$. A
by-product of maximizing the likelihood in (\ref
{eqnmarginallikelihoodalldata}) with the EM algorithm [detailed in
Supplementary Appendix~B of the supplemental article, \citet
{Kormaksson}] is posterior predictions of the latent indicators,
$(\mathbf{\hat{w}}_{c})_{c \in\mathcal{C}}$, which we round to
either~$0$ or~$1$. Given these estimated methylation indicators, the
conditional likelihood of a new observation $(y_{ij})_{j}$, on the
assumption that $i \in c$, is given by
%
%e10 ###
%e13 #&#
\begin{equation}
L_c(\boldtheta_i) = \prod_{j=1}^G \phi(y_{ij}|\mu_{1i},\sigma
_{1i}^2)^{\hat{w}_{cj}} \phi(y_{ij}|\mu_{2i},\sigma_{2i}^2)^{1-\hat
{w}_{cj}}. \label{discrimlik}
\end{equation}
The discriminant likelihood, $L_c$, is maximized with respect to the
individual specific parameters at
\begin{eqnarray*}
\hat{\mu}_{1i} &=& \frac{\sum_{j} \hat{w}_{cj} y_{ij}}{\sum_{j}
\hat{w}_{cj}},  \qquad  \hat{\sigma}^2_{1i} = \frac{\sum_{j} \hat
{w}_{cj}(y_{ij}-\hat{\mu}_{1i})^2}{\sum_{j} \hat{w}_{cj}},
\\
\hat{\mu}_{2i} &=& \frac{\sum_{j} (1-\hat{w}_{cj})y_{ij}}{\sum_{j}
(1-\hat{w}_{cj})},  \qquad
\hat{\sigma}^2_{2i} = \frac{\sum_{j} (1-\hat{w}_{cj})(y_{ij}-\hat
{\mu}_{2i})^2}{\sum_{j} (1-\hat{w}_{cj})}.
\end{eqnarray*}
By substituting these estimates into~(\ref{discrimlik}) we arrive at
the following discriminant rule:
%
%e11 ###
%e14 #&#
\begin{equation}
i \in c \qquad\mbox{if }   L_c (\hat
{\boldsymbol\theta}_i(\mathbf{\hat{w}}_c) ) > L_{c'} (\hat{\boldsymbol\theta}_i(\mathbf{\hat{w}}_{c'}) )
\mbox{ for all } c' \neq c. \label{equationdiscrimrule}
\end{equation}

%s6 ###
%s6 #&#
\section{Extension to multiple platforms} \label{secmultiplat}

In this section we discuss how to extend the methods of this paper to
account for multiple data types as long as each data type can
reasonably be modeled by the model described in Section \ref
{secmodelspecif}. For subject $i=1,\ldots,n$ let $y_{ijk}$ denote the
signal response of the $j$th variable, $j=1,\ldots,G_k$, on platform
$k=1,\ldots,m$. As before, we let $\mathcal{C}$ denote the true
partition of the $n$ subjects. We assume subjects in a given cluster $c
\in\mathcal{C}$ have identical activity (methylation, expression,
etc.) profiles on each platform $k=1,\ldots,m$ independently and define
a cluster and platform specific indicator for each variable
%
%e15 #&#
\[
w_{cjk} =
\cases{\displaystyle 1 ,&\quad  if variable $j$ on platform $k$ is
active in cluster $c$,\cr\displaystyle
0 ,&\quad  if variable $j$ on platform
$k$ is inactive in cluster $c$.
}
\]
Define $\mathbf{w}_c=(w_{cjk})_{j,k}$ and let $\mathbf
{y}_i=(y_{ijk})_{j,k}$ denote the vector of observed activity profiles
of subject $i$ across platforms. Let $\boldtheta_i=(\mu_{1ik},\sigma
_{1ik}^2,\mu_{2ik},\sigma_{2ik}^2)_{k=1}^m$ denote the subject
specific mixture parameters. We assume that the observed data, $\mathbf
{y} =(\mathbf{y}_1^T,\ldots,\mathbf{y}_n^T)^T$, given the unobserved
activity indicators, $\mathbf{w} =(\mathbf{w}_c)_{c \in\mathcal
{C}}$, arise from the following density:
%
%e16 #&#
\[
f(\mathbf{y}|\mathbf{w},\boldtheta) = \prod_{c \in\mathcal{C}}
\prod_{i \in c} f(\mathbf{y}_i|\mathbf{w}_c,\boldtheta_i),
\nonumber
\]
where the conditional density of $\mathbf{y}_i$, on the assumption
that $i \in c$, is given by
%
%e17 #&#
\[
f(\mathbf{y}_i|\mathbf{w}_c,\boldtheta_i) = \prod_{k=1}^m\prod
_{j=1}^{G_k} \phi(y_{ijk}|\mu_{1ik},\sigma_{1ik}^2)^{w_{cjk}} \phi
(y_{ijk}|\mu_{2ik},\sigma_{2ik}^2)^{1-w_{cjk}}. \nonumber
\]
We can either assume that the activity indicators for cluster $c$ are
fixed as in subection~\ref{seciterativeclustering}, or independent
Bernoullis, both across platforms and variables,
%
%e18 #&#
\[
f(\mathbf{w}_c) = \prod_{k=1}^m \prod_{j=1}^{G_k} \pi
_{1ck}^{w_{cjk}}\pi_{0ck}^{1-w_{cjk}}, \qquad\pi_{1ck} + \pi_{0ck} =
1, \nonumber
\]
where $\pi_{1ck}$ represents the proportions of variables on platform
$k$ that are active in cluster $c$. The likelihood in this integrated
framework is identical to the one given in (\ref
{eqnmarginallikelihoodalldata}), except we now have an additional
product across platforms $k$. The methods presented in Sections \ref
{sectionhierarchicalclustering} and~\ref{secclassification} thus
extend to multiple platforms in a straightforward manner.

%s7 ###
%s7 #&#
\section{Identifying subtypes of AML} \label{secanalysis}

Figueroa et~al. (\citeyear{Figueroa}) performed the first large-scale DNA methylation
profiling study in humans using the Erasmus data described in Section
\ref{secdatadescription}. They clustered patients using hierarchical
correlation based clustering on a subset of the most variable probe
sets. Using unsupervised clustering, they were able to classify the
patients into known and well-characterized subtypes as well as discover
novel clusters. In Section~\ref{secclusteringresults} we report
our clustering results on the data and compare to those of \citet
{Figueroa}. We ran a cluster analysis on both methylation and
expression data separately as well as an integrative cluster analysis
on both platforms simultaneously. In Section \ref
{secclassificationresults} we present results from a~discriminant
analysis study in which we classified an independent validation data
set using the methods of Section~\ref{secclassification}.

%s7.1 ###
%s7.1 #&#
\subsection{Clustering results} \label{secclusteringresults}

Figueroa et~al. (\citeyear{Figueroa}) hierarchically clustered the $n=344$ patients
(methylation profiles only) on a subset of the 3745 most variable
probe sets, using 1-correlation distance and Ward's agglomeration
method. These were probe sets that exceeded a standard deviation
threshold of $1$. We ran the hierarchical algorithm of Section \ref
{subsechierarch} on the same subset to obtain an initial partition.
Among the $344$ candidate partitions, obtained at each merging step,
the loglikelihood was maximized at $K=17$ clusters, but to avoid
singletons we chose a partition with $K=16$, the same number of
clusters \citet{Figueroa} chose. We then applied the iterative
algorithm of Section~\ref{seciterativeclustering} in an attempt to
improve upon the initial partition. We denote the resulting partition
``M'' (Methylation). We repeated this process separately for
the\vadjust{\goodbreak}
expression data using the 3370 most variable probe sets, or those that
exceeded a standard deviation threshold of $0.5$. This resulted in a
partition ``E'' (Expression) with $K=17$ clusters. Finally, we repeated
this process jointly on the 3745 and 3370 probe sets from the
methylation and expression data, respectively, resulting in the
partition ``ME'' with $K=14$ clusters.

\begin{figure}

\includegraphics{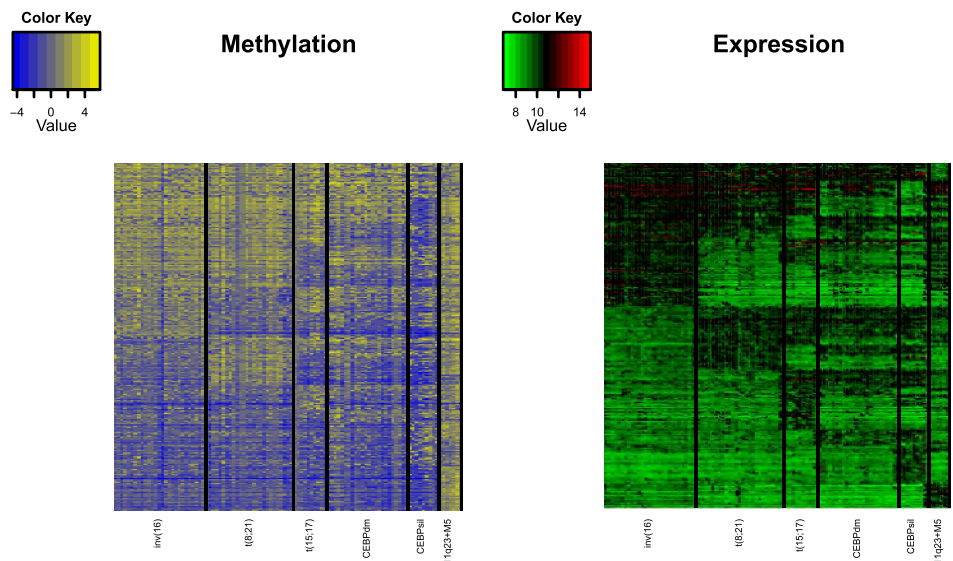}

\caption{Heatmaps of the methylation and gene expression values for
the 6 well-characterized clusters obtained from the integrative
clustering (``ME''). Columns correspond to patient samples and rows
correspond to genes. On the left heatmap ``yellow,'' ``gray'' and
``blue'' represent low, intermediate and high methylation,
respectively. On the right heatmap ``green,'' ``black'' and ``red''
represent low, intermediate and high expression, respectively. Rows are
ordered by increasing posterior probabilities of high
methylation/expression, $E[w_{cj}|\mathbf{y}]$, first for inv$(16)$,
then for $t(8;21)$, $t(15;17)$, CEBPdm, CEBPsil, and finally for
11q23+M5. Rows with equal probabilities for all $6$ clusters (either
equal to $1$ or $0$) are excluded to emphasize differences.}
\label{figheatmaps}
\end{figure}

\citet{Figueroa} identified $3$ robust and well-characterized
biological clusters and $8$ clusters that were associated with specific
genetic or epigenetic lesions. Five clusters seemed to share no known
biological features. The three robust clusters corresponded to cases
with inversions on chromosome $16$, inv$(16)$, and translocations
between chromosomes $15$ and $17$, $t(15;17)$, and chromosomes $8$ and
$21$, $t(8;21)$ (translocations refer to when genetic material from two
different chromosomes breaks apart and when being repaired, the
material from one chromosome is incorrectly attached to the other
chromosome instead and vice versa). The World Health Organization has
identified these subtypes of AML as indicative of favorable clinical
prognosis [see, e.g., \citet{Figueroa}]. The remaining $8$
clusters included patients with CEBPA double mutations (two different
abnormal changes in the genetic code of the CEBPA gene), CEBPA
mutations irrespective of type of mutation, silenced CEBPA (abnormal
loss of expression of CEBPA which is not due to mutations in the
genetic code), one cluster enriched for 11q23 abnormalities (any type
of change in the genetic code that affects position $23$ of the long
arm of chromosome $11$) and FAB M5 morphology (specific shape and
general aspect of the leukemic cell as defined by the French American
British classification system for Acute Leukemias), and four clusters
with NPM1 mutations (mutations in the genetic code of the NPM1 gene). A
detailed sensitivity and specificity analysis of $6$ of the above $11$
clusters [sample sizes in brackets], inv$(16)$ [$n_1=28$], $t(15;17)$
[$n_2=10$], $t(8;21)$ [$n_3=24$], CEBPA double mutations [$n_4=24$],
CEBPA silenced AMLs [$n_5=8$], and $11q23+\mbox{FAB M5}$ [$n_6=7$],
is given in Table~\ref{tablesensitivityJd} for the different
clustering results. We include the correlation based clustering result
(COR) on the methylation data to compare with ``M.'' The remaining five
of the $11$ biological clusters (CEBPA mutations irrespective of type
of mutation and the four NPM1 mutation clusters) had sensitivity or
specificity below $0.5$ for all four clustering results and were thus
excluded from the table. We can see that the model-based approach,
``M,'' is doing better than the correlation based method, ``COR,'' for
the most part. Aside for sensitivity of $t(8;21)$ ($1$ less false
negative) and specificity of inv$(16)$ ($1$ less false positive), the
model-based approach has as good or better sensitivity and specificity.
The most striking differences are in the numbers of false negatives of
CEBPA dm and false positives of $t(8;21)$ where ``M'' is doing better.
Note also that aside for the sensitivity of $11q23+\mbox{FAB M5}$
and specificity of $t(8;21)$, the integrated analysis ``ME'' always
does better than the analyses ``M'' and ``E'' separately, with perfect
sensitivity and specificity for many of the clusters. Most notably, the\vadjust{\goodbreak}
integrative analysis is able to perfectly classify the CEBPA double
mutants even though both ``M'' and ``E'' have quite a few false
positives and false negatives. This demonstrates the increased power to
identify clusters by sharing information across multiple platforms. As
a side product from our clustering algorithm, we obtain posterior
probabilities of high methylation/expression, $E[w_{cj}|\mathbf{y}]$,
which can be used to order genes in heatmaps to discover subtype
specific methylation/expression patterns. In Figure~\ref{figheatmaps}
we see heatmaps of the two data sets used for the integrative
clustering, ``ME,'' after rows have been ordered by increasing
posterior probabilities (one cluster at a time). Such heatmaps are
useful for graphically displaying the distinct methylation/expression
patterns that characterize the different subtypes of cancer.
%
%t1 ###
%t1 #&#
\begin{table}
\tabcolsep=0pt
\caption{The sensitivity and specificity of the clustering
results}\label{tablesensitivityJd}
\begin{tabular*}{\textwidth}{@{\extracolsep{\fill}}lcccc@{}}
\hline
\textbf{Subtype}&\textbf{COR}&\textbf{M}&\textbf{E}&\textbf{ME}\\
\hline
\multicolumn{5}{@{}c@{}}{Sensitivity (\# of false negatives in parentheses)}
\\
inv($16$) [$n_1=28$] & $0.929\ (2)$ & $0.964\ (1)$ & $0.857\ (4)$ &
$0.964\ (1)$ \\
$t(15;17)$ [$n_2=10$] & $0.800\ (2)$ & $0.800\ (2)$ & $1.000\ (0)$ &
$1.000\ (0)$ \\
$t(8;21)$ [$n_3=24$] & $0.917\ (2)$ & $0.875\ (3)$ & $0.917\ (2)$ &
$0.958\ (1)$ \\
CEBPA dm [$n_4=24$] & $0.792\ (5)$ & $0.917\ (2)$ & $0.75\ (6)$\hphantom{0}  & $1.000\ (0)$
\\
CEBPA Sil [$n_5=8$] & $0.625\ (3)$ & $0.875\ (1)$ & $1.000\ (0)$ & $1.000\ (0)$
\\
$11q23+\mbox{FAB M5}$ [$n_6=7$] & $0.857\ (1)$ & $0.857\ (1)$ &
$0.714\ (2)$ & $0.714\ (2)$ \\[4pt]
\multicolumn{5}{@{}c@{}}{Specificity (\# of false positives in parentheses)}
\\
inv($16$) [$n-n_1=316$] & $1.000\ (0)$ & $0.997\ (1)$ & $0.997\ (1)$ &
$1.000\ (0)$ \\
$t(15;17)$ [$n-n_2=334$] & $1.000\ (0)$ & $1.000\ (0)$ & $1.000\ (0)$ &
$1.000\ (0)$ \\
$t(8;21)$ [$n-n_3=320$] & $0.972\ (9)$ & $0.994\ (2)$ & $1.000\ (0)$ &
$0.991\ (3)$ \\
CEBPA dm [$n-n_4=320$] & $0.988\ (4)$ & $0.988\ (4)$ & $0.991\ (3)$ &
$1.000\ (0)$ \\
CEBPA Sil [$n-n_5=336$] & $1.000\ (1)$ & $0.997\ (1)$ & $0.985\ (5)$ &
$0.997\ (1)$ \\
$11q23+\mbox{FAB M5}$ [$n-n_6=337$] & $0.991\ (3)$ & $0.991\ (3)$ & $
0.991\ (3)$ & $ 0.994\ (2)$ \\
% Double horizontal line:
\hline
\end{tabular*}
\end{table}
%
%f2 ###
%f2 #&#

%s7.2 ###
%s7.2 #&#
\subsection{Classification results} \label{secclassificationresults}

A second cohort of patients with AML was available with which we could
test the performance of the classification method of Section \ref
{secclassification}. This second cohort of $n=383$ cases consisted of
samples, obtained from patients enrolled in a clinical trial from the
Eastern Cooperative Oncology Group (ECOG) (Data are available at
\href{http://www.ncbi.nlm.nih.gov/geo/}{http://www.ncbi.}
\href{http://www.ncbi.nlm.nih.gov/geo/}{nlm.nih.gov/geo/}, accession number pending). These
patients were similar in characteristics to the Erasmus cohort, with
only one exception, all patients were younger than $60$ years of age.
Samples were processed in the same way as the Erasmus cohort, and their
methylation was used to blindly predict their molecular diagnosis.
Using the 3745 most variable probe sets and the clustering result
``M'' of the previous section, we fit the model (\ref
{eqnmarginallikelihoodalldata}) on the Erasmus cohort with the EM
algorithm. By using the posterior predictions of the methylation
indicators, we applied the discriminant rule (\ref
{equationdiscrimrule}) on each patient in the ECOG data set. Since
CEBPA and NPM1 mutation status have not yet been made available for
this cohort, only the performance for the prediction of the inv$(16)$,
$t(8;21)$, CEBPA silenced, $t(15;17)$ and $11q23 + \mbox{FAB M5}$
clusters could be tested. Inv$(16)$ cases were predicted with $100\%$
sensitivity and specificity. The predicted $t(8;21)$ cluster contained
$100\%$ of cases positive for this abnormality, and only one $t(8;21)$
case was misclassified to another cluster. Two cases, which had
previously been unrecognized as CEBPA silenced AMLs, were predicted by
the classification method. One of them was later confirmed to indeed
correspond to this molecular subtype by an alternative methylation
measurement method. Similarly, one case was believed to have been
misclassified as $t(15;17)$ since there were no molecular data
confirming the presence of the PML-RARA gene fusion (the abnormal
combination of the PML and RARA genes) resulting from this
translocation. However, it was later confirmed that both the morphology
and the immune diagnosis corresponded to that of an acute promyelocytic
leukemia with $t(15;17)$. Finally, the $11q23+\mbox{FAB M5}$ cluster
included $9$ of the $14$ patients in the cohort that met these two
criteria. There were also $5$ false positives, $3$ of them were M5
cases but did not have $11q23$ abnormalities, $1$ of them harbored the
$11q23$ abnormality but corresponded to an M1, and the remaining case
corresponded to an M4 case with a hyperdiploid karyotype. A summary of
these results is provided in Table~\ref{tableclassification}.

%
%t2 ###
%t2 #&#
\begin{table}
\tablewidth=230pt
\caption{The sensitivity and specificity of the
classification result. False negatives and false positives are in
parentheses}\label{tableclassification}
\begin{tabular}{@{}lcc@{}}
\hline
\textbf{Subtype} & \textbf{Sensitivity} & \textbf{Specificity} \\
\hline
inv($16$) [$n=32$] & $1.000\ (0)$ & $1.000\ (0)$ \\
$t(15;17)$ [$n=1$] & $1.000\ (0)$ & $1.000\ (0)$ \\
$t(8;21)$ [$n=28$] & $0.964\ (1)$ & $1.000\ (0)$ \\
CEBPA Sil [$n=1$] & $1.000\ (0)$ & $0.997\ (1)$ \\
$11q23+\mbox{FAB M5}$ [$n=14$] & $0.643\ (5)$ & $0.986\ (5)$ \\
\hline
\end{tabular}
\end{table}

%s8 ###
%s8 #&#
\section{Other applications} \label{secotherapp}

The clustering method presented in this paper is not restricted to the
microarray platforms that the AML samples were processed on. In this
section we demonstrate the versatility of our method by applying it to
other microarray platforms and show that our algorithm does well in
clustering subjects. We also provide a comparison with other existing
methods for clustering microarray data.

%s8.1 ###
%s8.1 #&#
\subsection{Expression in endometrial cancer} \label{subsecendometrial}

In this subsection we analyze the microarray expression data set in
\citet{Tadesse}. Endometrioid endometrial
adenocarcinoma is a gynecologic malignancy typically occurring in
postmenopausal women. Identifying distinct subtypes based on common
patterns of gene expression is an important problem, as different
clinicopathologic groups may respond differently to therapy. Such
subclassification may lead to discoveries of important biomarkers that
could become targets for therapeutic intervention and improved
diagnosis. High density microarrays (Affymetrix Hu6800 chips) were used
to study expression of $4$ normal and $10$ endometrioid adenocarcinomas
on 7070 probe sets. Probe sets with at least one unreliable reading
(limits of reliable detection were set to 20 and 16,000) were removed
from the analysis, which resulted in $G=762$ variables. Finally, the
data were log-transformed, however, unlike \citet{Tadesse}, we chose
not to rescale the variables by their range. More details about the
data set can be found in \citet{Mutter} and is publicly available at
\url{http://endometrium.org}.

\begin{figure}

\includegraphics{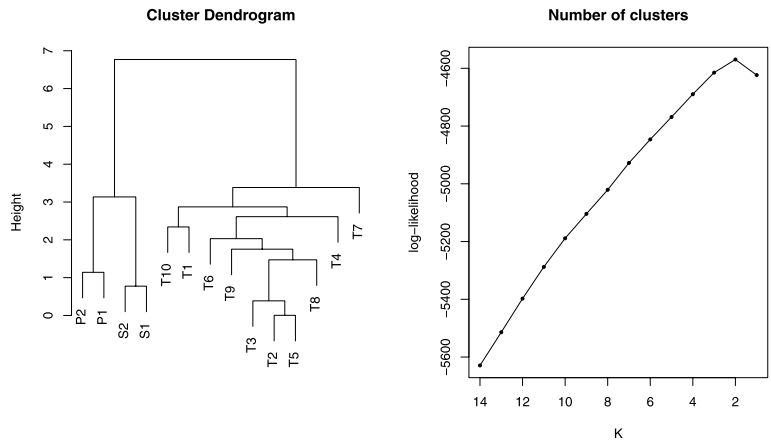}

\caption{Left panel: Cluster dendrogram for the $2$ normal
proliferative (``P'') endometria, $2$ normal secretory (``S'') endometria,
and $10$ endometrioid adenocarcinomas (``T''). Right panel: Plot of the
number of clusters against log-likelihood.}
\label{figendometryplot}
\end{figure}

We hierarchically clustered the samples using the $300$ most variable
probe sets and plotted the results in Figure~\ref{figendometryplot}.
We successfully separated the four normal tissues from the endometrial
cancer tissues and the log-likelihood plot suggests that $K=2$. The
dendrogram consistently separated the normals and\vadjust{\goodbreak} the cancer into two
branches for different variance thresholds. However, if we lowered the
threshold too much, the loglikelihood was maximized at $K=1$. This
makes sense, as including many low variability probe sets might mask
the true clustering structure. For comparison, \citet{Tadesse}
concluded that $K=3$ and commented that there could possibly be $2$
subtypes of endometrial cancer. However, to the best of our knowledge,
this subclustering has not been verified. They also reported clustering
results using the COSA algorithm of \citet{Friedman}, which, like our
analysis, seemed more suggestive of $K=2$.

%
%f3 ###
%f3 #&#

%s8.2 ###
%s8.2 #&#
\subsection{Methylation in normal tissues} \label{subsecnormaltissue}
Houseman et~al. (\citeyear{Houseman}) used the RPMM algorithm to cluster a methylation data
set consisting of $217$ normal tissues and compared the performance to
that of the HOPACH algorithm of \citet{Laan}. The RPMM analysis was
discussed in more detail in \citet{Christensen} and the data made
publicly available at the GEO depository with accession number GSE19434
(\url{http://www.ncbi.nlm.nih.gov/geo/}). Briefly, DNA was extracted from the
tissues, modified with sodium bisulfite, and processed on the Illumina
GoldenGate methylation platform. Average fluorescence for methylated
($M$) and unmethylated ($U$) alleles were derived from raw data at
$1505$ loci. However, in this study $1413$ loci passed the quality
assurance procedures. A total of $11$ tissue types were available,
bladder ($n=5$), adult blood ($n=30$), infant blood ($n=55$), brain
($n=12$), cervix ($n=3$), head and neck ($n=11$), kidney ($n=6$), lung
($n=53$), placenta ($n=19$), pleura ($n=18$) and small intestine
($n=5$). \citet{Houseman} constructed an average ``beta'' value from
raw data, which they claimed was very close to the quantity $M/(M+U)
\in(0,1)$, rendering a beta distributional\vadjust{\goodbreak} assumption (assuming~$M$
and $U$ follow a gamma distribution) with class and locus specific
parameters. We, however, chose to work with the quantity $\log(M/U)$,
which fits better with our two component mixture distributional
assumption. In order to get a direct comparison with RPMM and HOPACH,
we used all $1413$ loci. In Supplementary Figure 3 of the supplemental
article [\citet{Kormaksson}] we see a plot of cluster number versus
loglikelihood, which is maximized at $K=6$ clusters. However, given the
relatively small difference between the loglikelihood values at $K=6$,
$K=7$, and $K=8$, one could argue that all three clustering results are
worthy of consideration. The clusters are cross-classified with tissue
type in Table~\ref{tablecrossclassification}.

%
%t3 ###
%t3 #&#
\begin{table}
\tabcolsep=0pt
\caption{Cross-classification of our clustering result
($K=8$) and tissue type. By merging the top two clusters, and clusters
$3$ and $4,$ we obtain the clustering corresponding to $K=6$}\label
{tablecrossclassification}
\begin{tabular*}{\textwidth}{@{\extracolsep{\fill}}lccccccccccc@{}}
\hline
\textbf{Class}&\textbf{Blad}&\textbf{Bl}&\textbf{Br}&\textbf{Cerv}&\textbf{Inf bl}&\textbf{HN}&\textbf{Kid}&\textbf{Lung}&\textbf{Plac}&\textbf{Pleu}
&\textbf{Sm int}\\
\hline
1 & 5 & & & 2 & & 1 & & 53 & & 18 & 4 \\
2 & & & & 1 & & & & & & & \\
[5pt]
3 & & & & & & & 6 & & & & \\
4 & & & & & & 10 & & & & & 1 \\
[5pt]
5 & & 30 & & & & & & & & & \\
6 & & & 12 & & & & & & & & \\
7 & & & & & 55 & & & & & & \\
8 & & & & & & & & & 19 & & \\
\hline
\end{tabular*}
\end{table}

If we compare these results with those obtained with the RPMM
algorithm, our result favors few and concise clusters, whereas RPMM is
indicative of a total of $23$ subclasses of tissues. The HOPACH
clustering algorithm was suggestive of $K=9$ clusters, with $3$ of
those clusters representing placenta singletons separated from the main
placenta cluster. We present a~cross-classification table for both RPMM
and HOPACH in Supplementary Table 3 of the supplemental article [\citet
{Kormaksson}] for comparison [borrowed from \citet{Houseman}]. Our
method perfectly classifies blood, brain, infant blood, kidney and
placenta. For comparison, after bundling subclusters together, RPMM
classifies blood and infant blood perfectly, and HOPACH classifies
infant blood and placenta perfectly. All three methods have problems
distinguishing between bladder, cervical, lung, pleural and small
intestine tissues. Overall, our approach seems to outperform HOPACH,
and although \citet{Houseman} have demonstrated that a~few of their
tissue-specific subclusters (obtained by RPMM) have verifiable
meanings, such as through age difference, it seems that without a
further justification of such finer substructure in the data our
clustering result is more desirable. As a side note, under the
assumption of \citet{Houseman} that $M$ and $U$ follow a gamma
distribution, it is clear that\vadjust{\goodbreak} $\log(M/U)$ will not be a mixture of
two Gaussian distributions. The favorable clustering result for this
data set suggests that the normality assumption on each mixture
component provides a robust and flexible modeling distribution.

%s9 ###
%s9 #&#
\section{Discussion} \label{secdiscussion}

We have proposed a model-based method for clustering microarray data.
The methods have been demonstrated to work well on expression data and
methylation data separately. An integrated cluster analysis has further
shown the power of combining platforms in a joint analysis. We believe
this method can be applied to a variety of microarray data types.
However, further research is needed to validate the method on different
types of data such as ChIP--chip data.

A minor drawback of our method is that it does not allow for automated
selection of variables, but rather relies on pre-filtering the data.
However, most biologists are still relying on simple clustering
algorithms such as $K$-means or standard agglomerative algorithms, due
to their simplicity in implementation and interpretation. Thus, having
a relatively simple and easily implemented hierarchical algorithm that
can integrate multiple platforms and further utilizes the bimodal or
skewed structure of the individual profiles, in a model-based manner,
has its advantages. For example, the hierarchical algorithm
automatically determines the numbers of clusters and provides an easily
interpretable dendrogram. Also, as a side product, we obtain posterior
probabilities of high methylation/expression, $E[w_{cj}|\mathbf{y}]$,
for each cluster $c$ and probe set $j$. By ordering the probe sets with
respect to these posterior probabilities and excluding probe sets that
are identical across all clusters, we can explore patterns in heatmaps
such as in Figure~\ref{figheatmaps}.

One of the novelties of our clustering algorithm is the inclusion of
individual specific parameters, $(\mu_{1i},\sigma_{1i}^2,\mu
_{2i},\sigma_{2i}^2)_i$, into the model of Section \ref
{secmodelspecif}, which facilitates the use of our algorithm even in
the presence of extreme microarray effects. Since the amount of data we
have to estimate these parameters ($G$ observations per subject) highly
exceeds the number of subjects~($n$), the estimation of these
parameters has not been a problem. However, it is common practice to
treat such individual specific parameters as random effects. We have
established that with conjugate normal and inverse gamma priors on the
above parameters we arrive at a marginal likelihood intractable for
maximization. However, we have verified that through such prior
specifications we could easily calculate full conditionals in a
Bayesian analysis. A~Bayesian approach would also prevent us from
having to assume the methylation indicators, $\mathbf{w}$, are fixed
as in the iterative algorithm of Section \ref
{seciterativeclustering}. The reason for that assumption was that the
joint posterior distribution of $(\mathbf{w},\mathbf{X})$ is highly
intractable. However, full conditionals for each variable separately
are easily obtained and are that of Bernoulli and Multinomial,
respectively. Running a fully Bayesian analysis might also facilitate
an extended algorithm that could include all variables. One might
assume that some variables are informative and follow the mixture in
(\ref{eqnmarginallikelihoodalldata}) with prior probability $p$,
whereas other variables are noninformative with prior probability
$(1-p)$ and follow the mixture in (\ref
{eqnmarginallikelihoodalldata}) with $\mathcal{C}=[n]$. There are
some challenges that arise in implementing such a fully Bayesian model
and those require further research.

\section*{Acknowledgments}
The methods presented in this paper have been imple\-mented as an
R-package that
is available at \href{http://www.stat.cornell.edu/imac/}{http://www.stat.cornell.edu/}
\href{http://www.stat.cornell.edu/imac/}{imac/}.

\begin{supplement}%[id=suppA]
\stitle{Simulation and details of EM algorithms}
\slink[doi]{10.1214/11-AOAS533SUPP} %[doi,text={...}] - jei reikia
%suskaldyti doi
\slink[url]{http://lib.stat.cmu.edu/aoas/533/supplement.pdf}
\sdatatype{.pdf}
\sdescription{We perform a simulation study to assess the performance
of our clustering algorithm in the
presence of sparse correlation structure. We also derive the steps
involved in
maximizing the likelihoods of the several models presented in this paper.}
\end{supplement}

% imsref loaded by smiklovaite, 2012-02-05 09:25:31
%

\printaddresses

\end{document}